\documentclass[aps,prb,twocolumn,groupedaddress]{revtex4-2}

\usepackage{graphicx}
\usepackage{color}
\usepackage{amsmath}
\usepackage{enumitem}
\usepackage{amssymb}
\usepackage{hyperref}
\usepackage{cancel}
\usepackage{ulem}%%%%%%%

\usepackage{multirow}

\newcommand{\be}{\begin{equation}}
	\newcommand{\ee}{\end{equation}}

\newcommand{\bea}{\begin{eqnarray}}
	\newcommand{\eea}{\end{eqnarray}}

\newcommand{\p}{\partial}

\newcommand{\lb}{\left[}
\newcommand{\rb}{\right]}
\newcommand{\lp}{\left(}
\newcommand{\rp}{\right)}

\renewcommand{\vec}[1]{{\boldsymbol #1}}

\begin{document}
	
\title{Isospin- and momentum-polarized orders in bilayer graphene}
\author{Zhiyu Dong, Margarita Davydova, Olumakinde Ogunnaike, Leonid Levitov} 
\affiliation{Massachusetts Institute of Technology, Cambridge, Massachusetts 02139, USA} 
\date{December 25, 2022}

\begin{abstract}
Electron bands in the untwisted bilayer graphene flatten out in a transverse electric field, offering a promising platform for correlated electron physics. We predict that the spin/valley isospin magnetism, resembling that seen in moire bands, coexists with momentum-polarized phases occurring via a ``flocking transition'' in momentum space in which the electron distribution is spontaneously displaced in momentum space relative to the $K$ or $K’$ valley centers. These phases feature unusual observables such as persistent currents in the ground state.  Momentum-polarized carriers ``sample'' the Berry curvature of the conduction band, resulting in a unique behavior of the anomalous Hall conductivity and other effects that do not occur in previously studied systems.
\end{abstract}
\maketitle 

\section{Introduction}

Narrow bands in moir\'{e} graphene\cite{Bistritzer,Mele,SuarezMorell,Lopes_dos_Santos} host a variety of strongly correlated phases with exotic properties that can be accessed by tuning external fields and carrier density\cite{Cao1,Cao2,Kang,Xie,Ochi,Seo,Wu1, Wu2,Sharpe,Serlin,Tschirhart,Dodaro, Liu, Kerelsky, Choi2019, Jiang, Cao_2020,Guinea,Isobe,Kozii,Xu,Lin}. These findings inspired investigations into the existence of other narrow-band systems with interesting properties.
Recently, two nontwisted graphene multilayers---Bernal-stacked bilayers and ABC trilayers---have been identified\cite{HZhou_TLG,HZhou_BLG} as systems showing cascades of ordered phases resembling those seen in moir\'{e} graphene\cite{Saito,Zondiner,Choi2021,Rozen,Pierce}. These systems feature electron bands with field-tunable bandgaps and 
 dispersion that flattens out quickly as the field increases. Carriers in these bands become nearly dispersionless at large fields, 
forming strongly interacting systems with interesting properties\cite{Cvetkovic,Lee, Stauber,Castro,Castro2,Min,Throckmorton,Jung,Chou}. These developments prompted questions about new symmetry breaking types and new orders achievable in these systems. 

\begin{figure}[!t]
	\centering
	\includegraphics[width=0.47\textwidth]{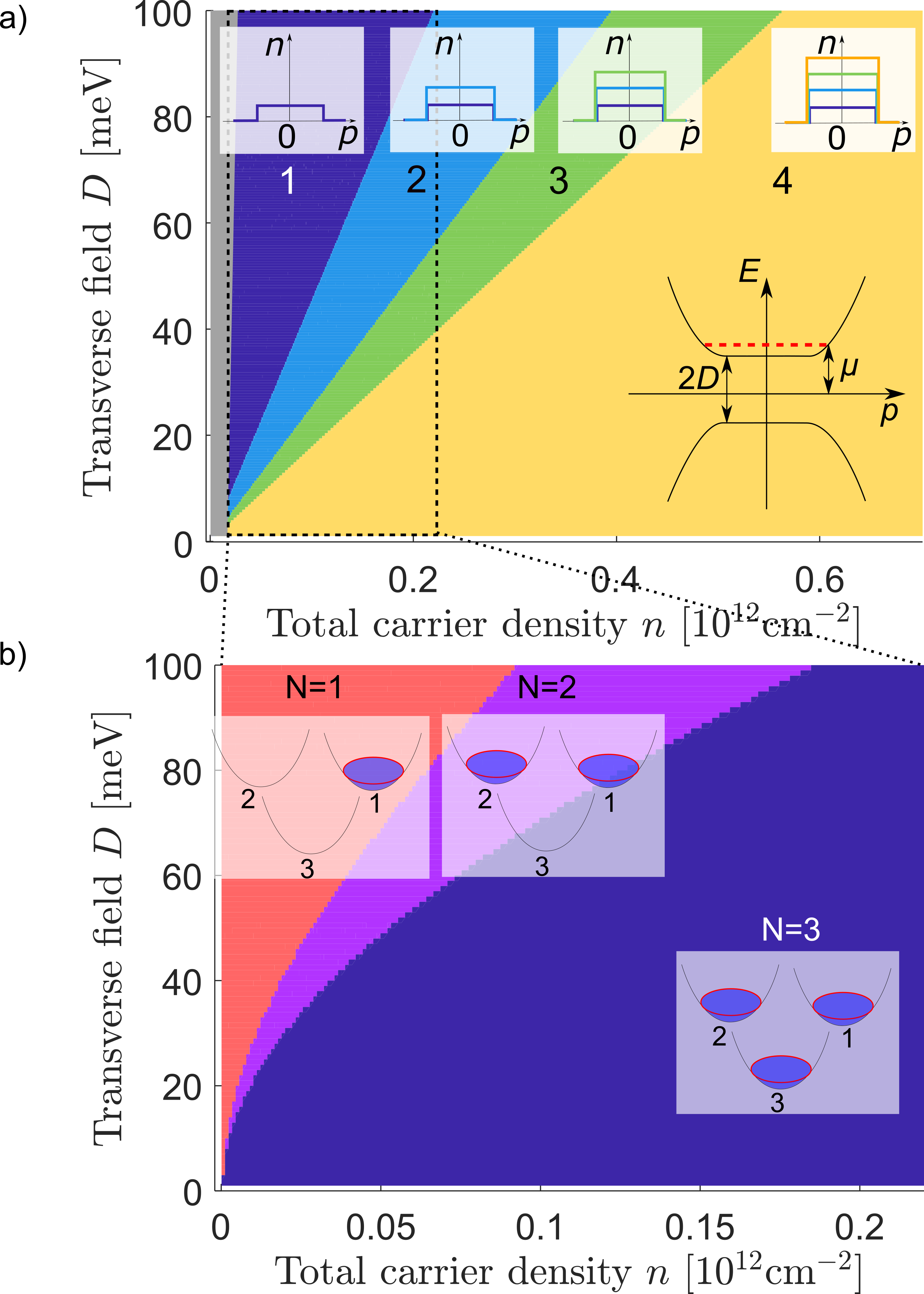}
	\caption{a) Phase diagram for different isospin (valley and spin) orders in a lightly-doped BBG band. Carriers form Fermi seas with the degree of isospin polarization increasing with field bias.  
	 States with different numbers of isospin species, pictured in the insets, are found in the four regions of the phase diagram obtained using realistic parameters. Polarization degree varies from 
	one layer (fully polarized) to 
	four layers (unpolarized) [see text beneath Eq.~\eqref{eq:n_D}]. b) Level-two symmetry breaking occurring in a dashed box marked in a). Different orders arise due to the Fermi sea spontaneously breaking into $N=1$, $2$ or $3$ pockets and shifting to different band minima [see Fig.~\ref{fig:pocket_model}].  
	}
	\vspace{-5mm}
	\label{fig:PD}
\end{figure}

Perhaps the most unusual aspect of these systems is that the flat band is not an isolated band, as in moir\'e graphene. Instead it is a flattened part of a dispersive band, with the degree of flatness and phase volume tunable by the displacement field.  
This behavior leads to properties distinct from those of unbiased bilayer graphene \cite{Nandkishore,CvetkovicBLG,Throckmorton}. 
Here, starting from a simple framework for the interaction effects in biased bilayer graphene (BBG), we predict new order types with properties considerably richer than those of isospin-ordered phases. %We focus on the flattened bands in field-biased bilayer graphene (BBG).%Bernal-stacked bilayer graphene (BG) induced by a transverse electric field. 
%Expectedly, strong interactions of carriers
Electron interactions in flattened bands drive isospin (spin-valley) polarization instability and a cascade of phase transitions between states with different polarizations, resembling those known in moir\'{e} graphene\cite{Saito,Zondiner,Choi2021,Rozen,Pierce}. A phase diagram for  
this cascade, derived below and shown in Fig.~\ref{fig:PD} a), strongly resembles the phase diagram seen experimentally\cite{HZhou_BLG,Sergio,Seiler}.

Further, an interesting change in behavior occurs at lower densities and stronger fields, where interactions lead to an isospin-polarized Fermi sea break-up and spontaneous momentum polarization, as illustrated in Fig.~\ref{fig:PD} b). Momentum-polarized phases originate from exchange-induced ``flocking'' effect, wherein all carriers  are shifted into one, two or three pockets at the band minima produced by the trigonal warping effects. These orders develop on top of the isospin-polarized phases. %\sout{that act as their mother states}.

\begin{figure}
	\centering
	\includegraphics[width=0.48\textwidth]{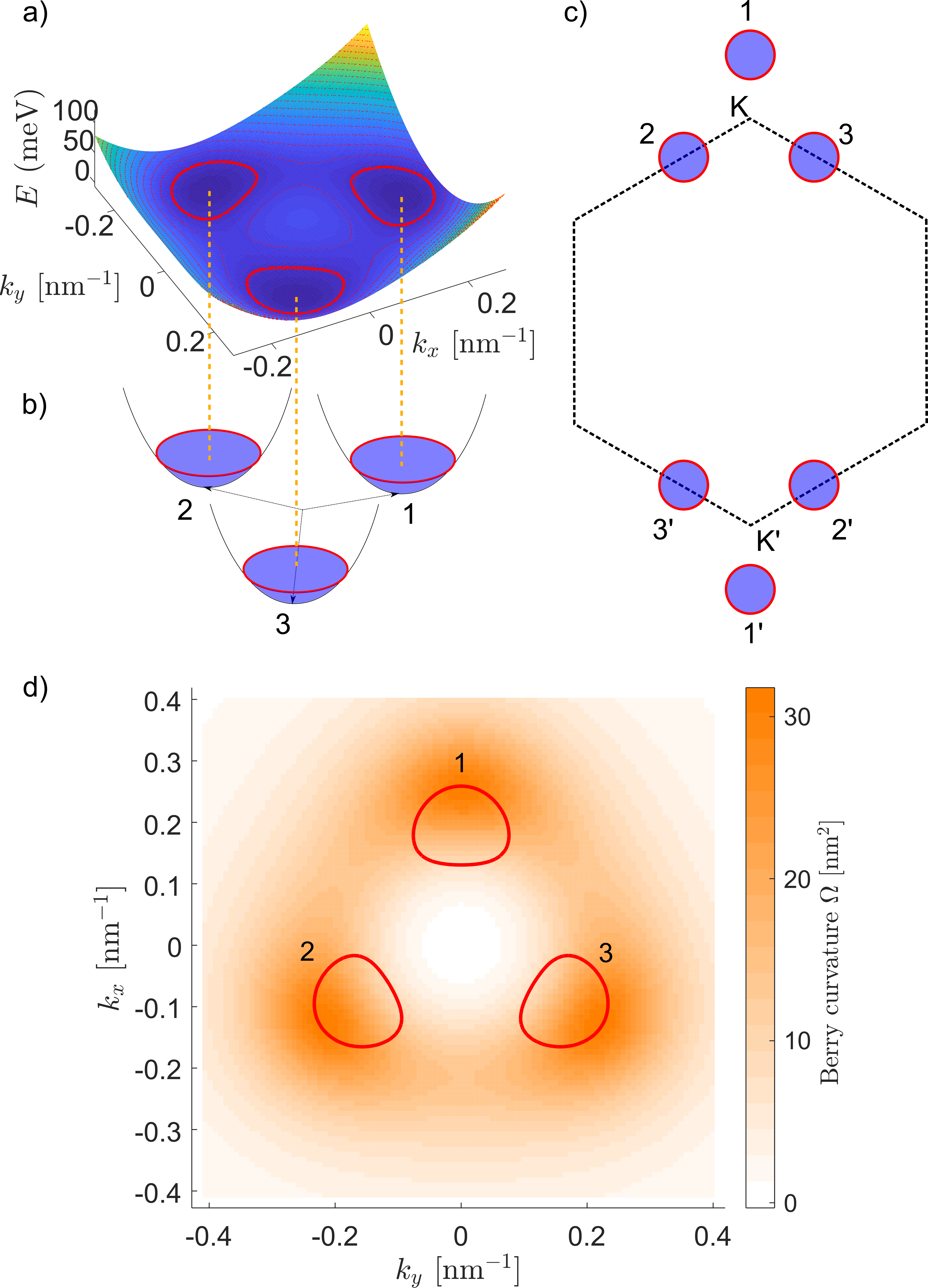}
	\caption{
		a) Conduction band dispersion flattened by transverse field. Trigonal warping interaction creates three mini-valleys, at low carrier density giving rise to three electron pockets (red contours). 
		b) A toy model for the three-pocket band structure. c) Schematic for pockets positioned near $K$ and $K'$ points. d) The distribution of the Berry curvature in the conduction band near $K$ point.  Parameters used: bias potential $D=100$ meV, chemical potential $\mu =90$ meV. The value $\mu<D$ reflects 
		the effect of the trigonal warping.
		}
	\label{fig:pocket_model}
	\vspace{-5mm}
\end{figure}

Momentum-polarized states are described phenomenologically by an effective Hamiltonian: 
\be
\tilde{H}(\vec k) = H(\vec k) + \vec u\cdot \vec k. \label{eq:uk}
\ee
Here $H(\vec k)$ is the trigonal warped Hamiltonian which we can approximate with three parabolic pockets displaced from $K$ points, as illustrated in Fig.~\ref{fig:pocket_model}, $\vec u$ is the order parameter that describes the polarization in momentum space.  As we will see, a spatial dependent $\vec u(\vec x)$ gives rise to a local persistent current and an orbital magnetization, which is allowed by the spontaneously broken time-reversal symmetry in momentum polarized phases. The persistent currents and orbital magnetization generated in this way are distinct from those familiar for the bands endowed with Berry curvature.

% We describe symmetry breaking transitions leading to these orders, and their unique signatures such as persistent current, electronic nematicity and incommensurate Kekul\'{e}-type density waves. 

%The spontaneously broken time-reversal symmetry of these phases results in 
%observables such as persistent currents and orbital magnetization that are distinct from those familiar for the bands endowed with Berry curvature. 
In addition to that, the momentum-polarized carriers “sample” the Berry curvature of the conduction band, leading to jumps and other unique signatures in the anomalous Hall effect. These effects are enhanced by the redistribution of Berry curvature 
throughout the conduction band resulting from its trigonal warping 
(see Fig.~\ref{fig:pocket_model}). 
An 
abrupt onset of a $B=0$ Hall effect, along with anisotropy of transport due to electronic nematicity, will provide clear signatures of momentum-polarized orders.

\section{An effective one-band model} % model for isospin polarization}

%\subsection{Two-band single-particle Hamiltonian}
Isospin polarization occurs when the exchange interaction between carriers in the conduction or valence band exceeds the kinetic energy. %This regime, arising
In the BBG system the kinetic energy is quenched when band dispersion is flattened by bandgap opening in the presence of a large transverse displacement field. This regime can be described by a one-band model derived by projecting the bilayer Hamiltonian to the conduction or valence band. Here we obtain this model starting from a conventional two-band model\cite{McCann}:
%
%with a relatively small carrier concentration such that the kinetic energy, reduced by the band flatness, is small compared to the exchange interaction. This regime can be described by an effective one-band model derived by projecting the bilayer Hamiltonian to the conduction or valence band. 
%
%In this section, we obtain a model for understanding the cascade of isospin transitions in BBG. The electrons in each valley of BBG is described by a two-band model\cite{McCann}:
%The two-layer model of graphene bilayer 
%The single-particle Hamiltonian can be written as:
%
\begin{align}\label{eq:H0}
	H_0 &= \sum_{\xi s \vec p}\Psi^\dagger_{\xi s \vec p} \left [h_0^{\xi} + h_t^{\xi}  + h_a +h_{D'}\right ] \Psi_{\xi s \vec p}
	\\ \nonumber
	h_0^{\xi} &= h_1(\vec p) \sigma_1 + h_2(\vec p) \xi\sigma_2 + D \sigma_3   \\ \nonumber
	%h_0 &= h_1(\vec p) \Sigma_1 + h_2(\vec p) \Sigma_2 + D \sigma_3   \\ \nonumber
	h_t^{\xi} &= v_3 \left ( p_2 \xi\sigma_1 + p_1 \sigma_2\right),\quad
	%h_t &= v_3 \left ( p_2 \Lambda_1 + p_1 \Lambda_2\right),\quad
	h_a = \frac{p^2}{2 m_a} ,\quad 
	h_{D'} = - D \frac{p^2}{\tilde p^2}  \sigma_3
\end{align} 
%
%and the coefficients of $h_0$ can be written as:
%
where $\xi=K,K'$ represents valley $K$ and $K'$, $s=\uparrow,\downarrow$ represents spin-up and spin-down, $\Psi_{\xi s\vec p } = (\psi_{A\xi s\vec p}, \psi_{B\xi s\vec p})^{\rm T}$, $\sigma_{1,2,3}$ are the Pauli matrices acting on the valley and sublattice (layer) degrees of freedom, respectively. 
%For conciseness, we have suppressed the spin indices for now, as  the spin-orbit coupling is negligible. 
The quantity $D$ is the interlayer bias generated by the transverse electric field. Here $h_1$ and $h_2$ are given by
\begin{align}
	h_1(\vec p) &=  \frac{1}{2 m} \left (p_1^2 - p_2^2\right )\\
	h_2(\vec p) &=  \frac{1}{2 m} \left (2 p_1 p_2\right ).
\end{align}
%and the matrices $\Sigma$ and $\Lambda$ are:
%\begin{align}
%	\Sigma_1 =  \sigma_1, \ \Sigma_2 = \xi \sigma_2, \
%	\Lambda_1 =  \xi \sigma_1, \
%	\Lambda_2 = \sigma_2
%\end{align}
%where $\xi=\pm1$ represents valley $K$ and $K'$.   
This model is derived  under the assumption that intra- and interlayer hoppings (A1B1  and A2B1-type terms in the original Hamiltonian, which are 3.16 eV and 0.38 eV, respectively) are much larger than all other energy scales. Note that we rotated the basis by 90$^\circ$ with respect to Ref. \cite{McCann}. 
%Here $h_0$ is the minimal model Hamiltonian that we start our consideration from in Eq.~\eqref{eq:H0}. 
Here $h_t$ produces the trigonal warping; $h_a$ produces the particle-hole asymmetry and $h_{D'}$ is the momentum-dependent contribution that is proportional to the displacement field with $\tilde p \approx 0.058/a_{\rm CC}$ a %$D$-independent 
constant (see Table \ref{tab:pars}).  

In this and next section, since we focus on understanding the isospin orders in $SU(4)$-symmetric model, we ignore for now the subleading terms\cite{McCann}, such as trigonal warping. These terms govern subtle effects such as momentum polarization, which will be considered in Sec. \ref{sec:three pocket model} and Sec.\ref{sec:momentum_polarize_phase diagram}.

We measure the energies in meV and the momentum is made dimensionless by multiplying by the carbon-carbon atom distance $a_{\rm CC} = 1.46 $ \AA. The relevant system parameters are given in Table \ref{tab:pars}.

\begin{table}[t]
	\begin{tabular}{|c|l|c|l|}
		\hline
		parameter & value & parameter & value \\
		\hline
		a & 2.46 \AA &  $v$ &  1.1$\times 10^6$ m/s \\
		$\gamma_0$    & 3.16 eV  &   $m_a$ & 0.19 $m_e$   \\
		$\gamma_1$  &  0.381 eV &  $m$ & 0.028 $m_e$    \\
		$\gamma_3$   & 0.38 eV  &  $v_3$ &  1.3$\times 10^5$ m/s     \\
		%$\gamma_4$ & 0.14 eV &  $v_4$ & 4.8$\times 10^4$ m/s    \\
		$D$ & $0 - 100$ meV  &  $\tilde p a_{\rm CC}$ &  0.058  \\
		\hline
	\end{tabular}
	\caption{Parameters in the Hamiltonian computed based on values in Ref.~\cite{McCann}. The velocities are defined as $v_i = (\sqrt{3}/2)a \gamma_i$ ($\hbar=1$ throughout this paper). The BG band mass is defined as $m = \gamma_1/2 v^2\approx 0.028 m_{\rm e}$. %, and the anisotropy mass is determined from $1/2 m_a = 2 v v_4/\gamma_1 + \Delta' v^2/\gamma_1^2$, where $\Delta=22$\,meV. 
	}  \label{tab:pars}
\end{table}

%\subsection{Projection on the conduction band}

%The projection to conduction band discussed in the main text can be formally obtained through the  transform %, which is valid since large D guarantees the upper band is isolated from lower band . Namely, we project all 4-by-4 matrices in Hamiltonian by
Here, we will be interested in the regime where the field-induced bandgap $2D$ is large compared to the carrier kinetic energy (see inset in Fig.~\ref{fig:PD} a)).  In this regime the upper and lower bands flatten out and effectively decouple. We therefore project the problem onto the conduction band %\cite{Supplement}
\be
%H_0\rightarrow 
\tilde{H}_0 = \hat{P}H_0\hat{P},
\ee
%where ${\rm tr}$ is the trace over sublattice degrees of freedom, 
Here, the projection operator $\hat P$ is defined as
\be
\hat P = \sum_\xi \frac{1}{2} \lp \frac{h_0^\xi (\vec p)}{E(\vec p)}+1 \rp, \quad E(p) =  \sqrt{D^2+ \lp \frac{p^2}{2m}\rp^2}.
\ee 
%After projection, %the Hamiltonian becomes
%\be
%\tilde{H} = \tilde{H}_0 + \tilde{H}_{int}
%\ee
%in which the single-particle part is
%the single-particle part of the minimal model Hamiltonian becomes diagonal
This yields a one-band Hamiltonian $\tilde{H}_0$ of the following form:
\be
\tilde{H}_0 = \sum_{i\vec p} E(\vec p) \tilde{\psi}_{i,\vec p}^\dagger \tilde{\psi}_{i,\vec p}, \quad i=K\uparrow, K\downarrow, K'\uparrow, K'\downarrow \label{eq:H02}
% \lp \tilde{\psi}_{pK}^\dagger \tilde{\psi}_{pK} + \tilde{\psi}_{pK'}^\dagger  \tilde{\psi}_{pK'} \rp
\ee
%\bea
%&& \tilde{H}_0 = \sum_p \tilde{h}_0(p) \tilde{\Psi}_p^\dagger \tilde{\Psi}_p \\
%&& \tilde{h}_0(p) =  1_\tau 
%\eea
where  %$i = K\uparrow, K\downarrow, K'\uparrow, K'\downarrow$, 
$\tilde{\psi}_{i,p}$ % = (\tilde{\psi}_{Kp},\tilde{\psi}_{K'p})^{T}$,
%is a four-component spinor 
is the field operator of conduction band electrons in two valleys and two spins. From now on, we write the spin indices explicitly. 

%In this paper, we focus on the effects of electron-electron interaction, which
Next, we modeled the electron-electron interaction as a density-density coupling:
\be
H_{int} = \frac{1}{2} \sum_{\vec p \vec p' \vec q}V_{\vec q} \psi^{\dagger}_{\alpha \xi s,\vec p} \psi^{\dagger}_{\beta \xi's' \vec p'} \psi_{\beta \xi's'(\vec p'-\vec q)}\psi_{\alpha \xi s (\vec p +\vec q)}, \label{eq:H_int1}
\ee
where $i$, $j$ and $\alpha$, $\beta$ label the isospin (spin-valley) and 	sublattice degrees of freedom, respectively.
At large $D$, the form of density-density interaction is invariant under projection:
\be
\tilde{H}_{int} = \frac{1}{2} \sum_{\vec p \vec p' \vec q}V_{\vec q} \tilde{\psi}^{\dagger}_{i,\vec p} \tilde{\psi}^{\dagger}_{j,\vec p'} \tilde{\psi}_{j,\vec p'-\vec q}\tilde{\psi}_{i,\vec p +\vec q } \label{eq:H_int2}
\ee
Similar to Eq.~\eqref{eq:H02}, subscripts $i$ and $j $ take values $ K\uparrow$,  $K\downarrow$, $K'\uparrow$, $K'\downarrow$. Here we ignore the intervalley Coulomb scattering because the interaction $V_{\vec q}$ drops as $1/q$, leading to intervalley interactions that are smaller than the intra-valley interaction by a factor of $p_F/2 K$ which is as small as $0.04$ at carrier density $10^{12}$cm$^{-2}$. 
As a result, the Hamiltonian, Eq.\eqref{eq:H_int2}, has an approximate isospin $SU(4)$ symmetry. We note parenthetically that at small $D$ field, Eq.~\eqref{eq:H_int2} is a no longer a good approximation since it should include a coherence factor which is depends on valley and momenta. This makes the strength of coupling between charge densities in valley $K$ different from the coupling between charge densities in valley $K$ and valley $K'$, breaking the approximate $SU(4)$ symmetry. However, this $SU(4)$-symmetry-breaking effect is small in the regime of interest. Namely, since this coherence factors are $1 - \mathcal O(E_F/D)$, the difference between the coherence factor in the intervalley density-density coupling term $\psi_K^{\dagger}\psi_K\psi_{K'}^{\dagger}\psi_{K'}$ and the one in the intravalley density-density coupling term $\psi_K^{\dagger}\psi_K\psi_{K}^{\dagger}\psi_{K}$ is at most $\mathcal O(E_F/D)$, which is small when displacement field $D$ is much larger than the Fermi energy $E_F$.

To gain insight into the parameter regime for isospin polarization occurs we use a simple constant interaction model, refining it in the subsequent analysis of momentum polarized order. 
The isospin order is a result of a Stoner instability arising from the exchange energy, which can be written as
\be
E_{\rm ex} = - \frac{1}{2}\sum_{i\vec p\vec p'} V_{\vec q-\vec q'}n_{i\vec q} n_{i\vec q}, \quad n_{i\vec q}=\langle \tilde{\psi}^\dagger_{i,\vec q}\tilde{\psi}_{i,\vec q }\rangle %\quad i=K\uparrow, K\downarrow, K'\uparrow, K'\downarrow.
\ee
where $i$ indexes isospin 
as in Eq.~\eqref{eq:H_int2}. Below, for simplicity, we model the interaction as a local interaction, $V_{\vec q-\vec q'} =V$.

Perhaps the closest comparison to our analysis of BBG in the literature is the early work on Stoner spin instability in BBG \cite{Castro, Castro2,Stauber}. 
This work employed an atomic-scale short-ranged interaction, which did not allow treating valley and spin degrees of freedom on equal footing. 
The interaction used here, in contrast, is blind to valley and spin, leading to an approximate $SU(4)$ symmetry and a cascade of isospin orders.

\section{Broken isospin $SU(4)$ symmetry and phase diagram}%Isospin polarization}

%\addQ{Using the $SU(4)$ symmetric model given in last section
Next, we proceed to analyze the isospin polarization orders using as a framework the $SU(4)$ symmetric model introduced above. The onset of $SU(4)$ isospin polarization is 
determined by the Stoner criterion:
\be
V\nu = 1,
\ee
with the density of states $\nu$ (per isospin species) in the conduction band,
\be
\nu = \frac{m}{2\pi}\frac{\mu}{\sqrt{\mu^2-D^2}} \approx \frac{m^2}{(2\pi)^2}\frac{D}{n}.
\ee
Here we have used the expression for the electron density in the single-electron picture, $n = \frac{m}{2\pi} \sqrt{\mu^2-D^2}$, taking the chemical potential to lie near the bottom of the band,  $\mu\approx D$. With this, we estimate the carrier density at the onset of the Stoner instability, finding a fan of phase boundaries $n$ vs. $D$ for $M=1$, $2$ or $3$ isospin species:
\be
n_{D} =  M\frac{V m^2 D}{(2\pi)^2} \label{eq:n_D}
.
\ee
While in general the mean field Stoner approach has limitations, in this case it appears to be accurate. For BBG parameters $m=0.028 m_{\rm e}$\cite{McCann,Mayorov,Velasco}, $V = 10^3$\,meV\,nm$^2$ [see Appendix \ref{sec:estimate V} ], this simple model predicts an isospin ordering transition at carrier densities $n_D \sim 10^{12}$cm$^{-2}$ for the interlayer bias $D=100$ meV, in excellent agreement with Ref.\cite{HZhou_BLG}. 

The mean-field phase diagram, obtained by comparing energies of partially polarized states with $M=1$, $2$, $3$ and $4$  
isospin species found numerically, is shown in Fig. ~\ref{fig:PD} a). 
The yellow area represents the disordered phase where all four isospin species are equally filled. Purple, light blue and green mark stability regions for isospin-ordered states. 
The inset in the lower right corner shows electron dispersion near charge neutrality, with the Fermi level marked by a red dashed line.  
The insets at the top illustrate the layer-cake structure of electron distribution in each of the phases, with  the Fermi seas for different isospin species shown in different colors.  
The gray region near charge neutrality marks the band insulator phase with an unoccupied conduction band. 
The dashed rectangle marks the region of low carrier density on which the second half of this paper will focus. 
As we will see in Sec.\ref{sec:three pocket model} and Sec.\ref{sec:momentum_polarize_phase diagram}, trigonal warping of the conduction band flattened by the external field $D$ gives rise to Fermi sea breakups and level-two symmetry breaking through spontaneous momentum polarization. This behavior is summarized in the phase diagram in Fig.~\ref{fig:PD} b).

\begin{table*}[htbp]
	\begin{tabular}{||c | c| c | c | c | c | c ||} 
		\hline
		irreps & matrices & $O_{1}$ &  broken symmetries & Ohmic conductivity & spatial modulation & Hall conductivity\\
		\hline
		%$A_{1,\Gamma}$ & - & - &  -& \\ 
		%\hline
		$A_{2,\Gamma}$, $1$D & $\tau_3$ & $ O_1^{z} =P^{\pm z}_{\tau}P^{\vec m}_{s} $ & mirror, time reversal &  isotropic & none & nonvanishing \\ 
		\hline
		%$E_{2,\Gamma}$ & - &- & rotation, mirror & \\ 
		%\hline
		%$A_{\pm K}$ & - &- &  mirror, translation&  \\
		%\hline
		$E_{\pm K}$, $2$D & $\lp \tau_{1}, \tau_{2} \rp $ & $O_1^{xy} = P^{\vec \gamma}_{\tau} P^{\vec m}_{s}$ & rotation, mirror, translation& anisotropic & Kekul\'e order & vanishing \\
		\hline
	\end{tabular}
	\caption{Symmetry classification of different %momentum-independent 
	isospin orders. Listed are results for two real irreducible representations (irreps) of the BBG space group under which the valley-space Pauli matrices $\tau_{1,2,3}$ %(the second column) 
	can transform; other irreps are not realized by isospin-polarized orders. Column 1 lists the irreps and their dimensions. In column 3, the projection operators in valley and spin space constituting the order parameter are: $P^{\pm z}_{\tau} = \frac{1}{2}\lp 1 \pm \tau_{3} \rp$, $P^{\vec \gamma}_{\tau} = \frac{1}{2}\lp 1+\gamma_1 \tau_{1}+\gamma_2 \tau_{2} \rp$, $ P^{\vec m}_{s} = \frac{1}{2}\lp 1 + \vec s \cdot \vec m \rp$, where $\vec \gamma = (\gamma_1,\gamma_2)^{\rm T}$ with real $\gamma_{1,2}$, %is an arbitrary real vector \addQ{in $xy$ plane,} 
	$\vec m = (m_1,m_2,m_3)^{\rm T}$ is an arbitrary three-dimensional real vector. Columns 4-7 list broken symmetries and signature observables (see 
	text). 
		\label{table:symmetry}
		\vspace{-3mm}
	}
\end{table*}

Because of the $SU(4)$ symmetry of our model, the phase diagram in Fig.~\ref{fig:PD} a) is 
insensitive to the order parameter orientation in the isospin space. 
However, in reality, small valley anisotropy in the Hamiltonian, e.g. trigonal warping or intervalley scattering, can lift the $SU(4)$ degeneracy and favor a certain orientation in isospin space. Here, rather than analyzing the competition between phases with different symmetry, we take a general symmetry approach and list all possible way of breaking the $SU(4)$ symmetry. The energetics describing this competition will be discussed elsewhere. 

%The trigonal warping effects can also lead to additional symmetry breaking such as pocket polarization described in the next section. We ignore th

Our symmetry analysis benefits from the observation that the symmetry aspects of different orders and the general properties of the order parameter can be understood regardless of detailed knowledge of which order is ultimately favored. %ultimately wins. 
Below, we describe the possible order types, classify them through the symmetry of our problem. 
%These competing effects will be discussed elsewhere. 
%Yet, the symmetry aspects of different orders can be understood very generally without detailed knowledge of the order that is ultimately favored. %ultimately wins. 

For simplicity, %and since the effect of interaction is strongest at lowest density, 
we focus on the case of phase 1 where electrons only occupies one isospin species. %,  where only one of the isospin species is populated. 
Other orders can be studied in a similar manner.
Table~\ref{table:symmetry} summarizes the results %of our symmetry analysis%\cite{Supplement} 
for phase 1. In this case there are two possible phases, $O_1^{z}$ and $O_1^{xy}$, describing orders with valley imbalance and intervalley coherence, respectively. These two order types break different symmetries and have different signature observables as a result. %\cite{Supplement}. 

%-------------------

%However, the symmetry of all possible orders is one important aspect of the system and is interesting to analyze in detail. 
We arrive at this  conclusion as follows. 
 In phase 1 the order parameter is simply a projection onto the state with a given valley-spin orientation. Therefore, it takes the form of 
\be
O_{1}= |v \rangle \langle v|
\ee
where $v$ is an arbitrary normalized complex-valued four-component spinor in the isospin space, $|v \rangle = (\alpha_1 |u_1\rangle,\alpha_2 |u_2\rangle )^{\rm T}$ where $|u_1\rangle, |u_2\rangle$ are arbitrary normalized two-component state vectors in the spin subspace, $\alpha_1,\alpha_2$ are positive real numbers, $\alpha_1^2+\alpha_2^2 = 1$. Overall phases are absorbed in $|u_1\rangle$ and $|u_2\rangle$. The symmetry analysis of the Pauli matrices in valley basis (see Table \ref{table:symmetry}) indicates that  $\tau_{1,2}$ and $\tau_3$ transform under different irreducible representations. Thus, an order parameter containing $\tau_{1,2}$ matrices and another one containing $\tau_3$ corresponds to different broken symmetries. Therefore, to classify orders by symmetry, we look for an order parameter, $O$, that contains $\tau_3$ or $\tau_{1,2}$ matrices only, but not a mixture of $\tau_3$ and $\tau_{1,2}$. %This gives two types of solutions with different kind of valley order, whereas the orientation of spin parametrized by $|u_{1,2}\rangle$ remains arbitrary. 
This gives two possible types of the order parameter with distinct symmetry: $O_{1}^{z} = \frac{1}{4}\lp 1\pm \tau_3 \rp \lp 1 + \vec s\cdot \vec m\rp $ and $O^{xy}_{1} = \frac{1}{4} \lp 1+ \gamma_1\tau_{1} + \gamma_2\tau_{2} \rp \lp 1 + \vec s \cdot \vec m\rp $. Here, $\vec m$ is an arbitrary vector determining the spin direction, and $(\gamma_1,\gamma_2)$ is an arbitrary normalized real-valued vector. 

The order $O^z_{1}$ represents a valley imbalance order, which transforms under $A_{2,\Gamma}$ and thus, features a breakdown of the mirror symmetry that swaps the two valleys. The %other  option
second order parameter, $O^{xy}_{1}$, corresponds to the intervalley coherent order that transforms under $E_{\pm K}$. It breaks the three-fold rotation, reflection and translation symmetries of the original model. This aspect clearly differentiates the AB bilayer graphene from the case of ABC trilayer: in the latter, the intervalley coherent state does not break the $C_3$ rotation symmetry\cite{Cvetkovic}. The symmetry classification of possible orders in AB bilayer graphene is summarized in Table.\ref{table:symmetry}. %in the main text. 

Our symmetry analysis allows us to identify two observables that distinguish the valley imbalance and intervalley coherent orders in phase 1. These are anisotropy of conductivity and a spatial charge density wave modulation. For valley imbalance order $O^{z}_{1}$, neither rotation nor translation symmetry of the space group is broken, so the conductivity is isotropic and there is no spatial pattern. In comparison, for the valley coherence order $O^{xy}_{1}$, both rotation and translation symmetries are broken. The broken rotation symmetry leads to an anisotropic conductivity, whereas the broken translation symmetry leads to a spatial pattern with momentum $2K$, i.e. a Kekul\'e charge density wave. 
On a different note, the temporal symmetry can be probed by the Hall conductance. For the valley imbalance order $O^{z}_{1}$ where time reversal symmetry is broken, the Hall conductivity is nonvanishing. In comparison, the intervalley coherent order $O^{xy}_{1}$ preserves time reversal symmetry, guaranteeing %which guarantees 
a vanishing Hall conductance.
These observables are summarized in the last three columns in Table~\ref{table:symmetry}. %in the main text.  

\section{Momentum-polarized order: the three-pocket model}\label{sec:three pocket model}

Next, we turn to discussing momentum-polarized ordered states that are unique %[reword? The referee does not like it]} 
to BBG. These orders are triggered by Lifshits transition in which an isospin-polarized Fermi sea splits into several distinct pockets centered around the minima of the conduction band. Following this transition, exchange interactions drive symmetry breaking between different pockets through a momentum polarization instability.
		
It is instructive to start with a qualitative discussion of how this instability comes into play. There is an anisotropy in a realistic BBG bandstructure at small momenta due to the trigonal warping term, which is not included in the minimal description of band structure Eq.~\eqref{eq:H0}. This anisotropy leads to a three-pocket shape of Fermi surface in the regime of extremely low carrier density. As a result, for each isospin, instead of uniformly filling all three pockets, there are three candidate electron configurations for the ground state, in which either one, two or all three pockets are filled. Which one wins is determined by the competition between the kinetic and the exchange energy. The kinetic energy favors the configuration where all pockets are uniformly filled, whereas the exchange energy is optimized when all electrons are placed in the same pocket, since the interpocket exchange interaction is weaker than the intrapocket one.

To estimate of the energy scales that govern this competition, we consider the total single-particle kinetic energy for all carriers polarized in one pocket: 
\be
E_{\rm kin} \sim n^2/2\nu_*
,
\ee
where $\nu_* \sim 5\times 10^{-5}$ meV$^{-1}$ nm$^{-2}$ is 
the density of states at the bottom of a single pocket, obtained using the pocket dispersion parameters estimated below. %\cite{Supplement}. 
To study the pocket order, we take into account the momentum dependence of the interaction. Then the exchange energy is:
\be\label{eq:pocket_exchange_energy}
E_{\rm ex} \sim -\frac{2\pi e^2  }{\kappa |p|} n^2 \sim -\frac{\sqrt{\pi}}{\kappa} e^2 n^{3/2}
,\quad 
|p|\sim \sqrt{4\pi n}
,
\ee
with the characteristic momentum scale estimated from particle spacing. 
As a result, the exchange energy dominates at sufficiently small density $n\lesssim n_* =  \frac{4\pi}{\kappa^2}e^4 \nu_* ^2 \sim 10^{11}$ cm$^{-2}$,  
where we have used a realistic value $D\sim 100$ meV
and the dielectric constant $\kappa\sim 5$ similar to dielectric constant in monolayer graphene. The realistic $\kappa$ values will depend on the experimental setup.

%\addZ{We formulate this problem using a toy model with $1/k$ Coulomb interaction\cite{Supplement} that leads to a generalized Stoner instability at small density, described by an effective Hamiltonian Eq.~\eqref{eq:uk}.}
%\be
%\tilde{H}(\vec k) = H(\vec k) + \vec u\cdot \vec k. \label{eq:uk}
%\ee
%Here $H(k)$ is the trigonal warped Hamiltonian which we can approximate with three parabolic pockets displaced from $K$ points, as illustrated in Fig.\ref{fig:pocket_model}. The order parameter $\vec u$ describes the polarization in momentum space. Its direction determines which one of the three pockets is filled.} 

The resulting %pocket polarization 
phase diagram in the small density regime is shown in Fig.~\ref{fig:PD} b). %Here, we only show the part of the phase diagram where our analysis is valid. %(see Eq.\eqref{eq:validity_n} and Eq.\eqref{eq:validity_D}) 
%From this phase diagram, we find that electrons always prefer to cluster in one of the three pockets at a low carrier density. 
At lowest carrier density, exchange energy dominates and all electrons prefer to  polarize in a single pocket. Upon carrier density increasing, the system undergoes phase transitions, first to a two-pocket configuration, and then to a three-pocket (unpolarized) phase. %Increasing the screening suppresses the single-pocket order since the single-pocket order is generated by the exchange interaction.
For illustration, in Fig.~\ref{fig:PD} b), we set the dielectric constant to be $\kappa = 3$, so that the phase diagram showcases all possible phases. The details of the phase diagram observed in experiment may vary from system to system, since the competition of pocket orders is sensitive to screening effects that %, and that 
depend on the experimental setup. If screening is made stronger [e.g. by a proximal gate], the pocket-ordered state will be suppressed compared to that shown in Fig.~\ref{fig:PD} b). Alternatively, if the screening is made weaker, the pocket-polarized phase will expand, taking over a larger part of the phase diagram.
	%For example, if screening is much stronger then the strength we choose in Fig.\ref{fig:pocket_model_PD} due to a proximal gate, then the pocket polarized state can be completely suppressed. If screening is much weaker than the value we chose in Fig.\ref{fig:pocket_model_PD}, then the pocket polarized order will take over the whole $1/4$ phase.} 
We note that the energy difference between the pocket polarized and unpolarized states is of the order of $0.1$meV to $1$meV per carrier, yielding a readily accessible ordering temperature scale of a few kelvin.

%In this section, we 

We end this section with detailing the procedure through which we extract the parameters $k_{*}$ and $m_{*}$ used in the three-pocket model by starting from the realistic BG band structure. As we only care about the band dispersion near the band minima, we model the three-pocket band structure using three isotropic parabolic bands:
\be
H_{\alpha}(\vec p) = \frac{(\vec p-\vec k_\alpha)^2}{2m_{*}}, \quad \alpha=1,2,3 \label{eq:toy_model_1}
\ee
Here $\alpha$ labels the pockets, $p_\alpha's$ are the centers of pockets, corresponding to three minima of the conduction band: 
\be
\vec k_1 = k_* (0,1),\quad \vec k_{2,3} =  k_*\lp \mp \frac{\sqrt{3}}{2},-\frac{1}{2}\rp.
%, \quad \vec k_3 = k_*(\frac{\sqrt{3}}{2},-\frac{1}{2}).
\ee 
The values of $k_*$ and $m_*$ will be specified below. 

In order to relate the three-pocket bandstructure represented by three parabolas to the single-particle bandstructure shown before, we first adopt a minimal model that possesses thee pockets at large displacement field $D$. This model has the Hamiltonian
\begin{align} \label{eq:Hmin}
H_{3-p}^{min} = \sum_{\vec p}\psi^\dagger_{i\vec p} \left [h_0(\vec p) + h_t(\vec p)  + h_{D'}(\vec p)\right ]_{ij} \psi_{j\vec p}
\end{align}
Here $i$ labels isospin, the term $h_{D'}$ is responsible for the Mexican-hat shape dispersion. %in the absence of trigonal warping $h_t$
We find the value of $k_*$  by neglecting the trigonal warping term: 
\be \label{eq:sup k_*(D)}
k_*(D) \approx \tilde p  \frac{D}{\tilde{E}}, \quad \tilde{E} = \frac{\tilde p^2}{2 m} \approx 0.2 \text{ eV}, %\quad k_* (D = 100 \text{ meV}) a_{\rm CC} \approx 0.03
\ee
The trigonal warping determines the positions of the three minima of the conduction band but has a negligible effect on the radial coordinate of these minima.

%In the main text, 
%We use the expression $k_*(D)$ given in Eq.\eqref{eq:sup k_*(D)} when numerically computing phase diagram. But when estimating the validity condition of toy model, we used the value of $k_*$ at $D=100$ meV for simplicity, which is $k_*(100$meV$)\approx 0.03/a_{\rm CC}$.

The mass $m_{*}$ is a parameter that we introduced in the three-pocket toy model to mimic the bottom of the conduction band from \eqref{eq:Hmin}. The Hamiltonian Eq.\eqref{eq:Hmin} near one of the minimum takes the following form
\be
H(\vec k_1 + \delta \vec p) = \frac{\delta p_x^2}{2m_{*\perp}} + \frac{\delta p_y^2}{2m_{*\parallel}},
\ee
We find that the effective mass in radial direction $m_{*\parallel}$ is determined mainly by $h_D$, whereas the perpendicular mass $m_{*\perp}$ is only finite when we include the trigonal warping term:
\be
m_{*\parallel} \approx  \frac{m \tilde{E} }{4D}, \quad m_{*\perp} \approx \frac {\tilde p}{6 v_3}.
\ee
%\addZ{$m_{*\parallel} = 5\times 10^{-6}/0.146^2$ meV$^{-1}$ nm$^{-2} \sim 2.3 \times 10^{-4}$}
%\addZ{$m_{*\perp} = 8.3\times 10^{-6}/0.146^2$ meV$^{-1}$ nm$^{-2} \sim 3.9 \times 10^{-4}$}
%As we are interested in instabilities, the most important quantity that we need to mimic with toy model is the density of states. Therefore, below, we determine the value of $m_*$ in our toy model so that it reproduces the density of states of the original model. The density of states in the model \eqref{eq:Hmin} is given by
%\be
%\nu = \frac{1}{2\pi} \sqrt{m_{*\perp}m_{*\parallel}}.
%\ee
In our three-pocket toy model Eq.\eqref{eq:toy_model_1}, we set the parameter $m_*$ as
\be
m_{*}=\sqrt{m_{*\perp}m_{*\parallel}}.	
\ee
so that the three-pocket model reproduces the density of states of the realistic band structure.

\section{Stoner instability in the pocket channel. Phase diagram.} \label{sec:momentum_polarize_phase diagram}
%Here we consider a simple model that accounts for pocket ordering and work out a phase diagram for realistic system parameters. We use a simple low-energy model consisting of three parabolic bands representing the three electron pockets in the conduction band: 
%\be
%H_{i}(\vec p) = (\vec p-\vec k_i)^2/2m_{*},  \label{eq:toy_model}
%\ee
%where $i=1,2,3$ labels the pockets, with the pocket centers $\vec k_i$ positioned at the three minima of the conduction band, 
%$\vec k_1 = k_* (0,1)$, $\vec k_{2,3} = k_*(\mp \sqrt{3}/2,-1/2)$. 
%The values of  $k_*$ and the effective masses $m_*$ of the pockets are extracted from a realistic band structure as discussed above. %\cite{Supplement}.
Using the three-pocket model, we proceed to analyze the instability toward momentum-polarized state and obtain a phase diagram. For clarity, we focus on the effects arising in phase 1 [see Fig.\ref{fig:PD}], where the additional effects of densities in different isospin states is absent. There are three possible candidate ground states in which electrons fill up one, two or all three pockets. To determine which one of them is the true ground state, we compare their energies $E_{N}$ ($N=1,2,3$ is the number of occupied pockets) at the same total carrier density $n$. Their energies $E_{N} = E_K^{(N)}  + E_{ex}^{(N)}$ consist of kinetic and exchange energy contributions.
Using the fact that the density of states in each pocket is a constant $\nu_* = m_*/2\pi$, we can write the total kinetic energy as
\be
E_{K}^{(N)}  = \frac{N}{2\nu_*} \frac{n^2}{N^2} = \frac{\pi n^2}{ N m_*}\label{eq:E_K_result} 
\ee
In order to explore the pocket polarization, we restore the momentum dependence of the interaction in the exchange part of the free energy:
\be
E_{ex}^{(N)} = -\frac{1}{2} \sum_{i,j=1}^N \sum_{\vec p,\vec p'} V_{p-p'}n_{i\vec p} n_{j\vec p'},  \quad V_{\vec p-\vec p'} = \frac{2\pi e^2}{\kappa |\vec p-\vec p'|}, \label{eq:exchange1}
\ee
where 
$n_{i\vec p}$ is the occupation number at momentum $\vec p$ measured 
relative to the pocket $i$ center.
For simplicity, as in Eq.\ref{eq:pocket_exchange_energy}, we use momentum-independent dielectric constant $\kappa$. 
When the carrier density is small, 
the interpocket exchange interactions yield a nearly momentum-independent renormalization of the energy of each electron, which justifies approximating the Fermi surfaces in the pockets by discs centered at $\vec k_{i}$. This yields an estimate for exchange energy %From this, we have
\begin{align}
E_{ex}^{(N)} %= -\frac{1}{2}\sum_{i,j=1}^N \sum_{\vec p,\vec p'} V_{\vec p-\vec p'}n_{i\vec p} n_{j\vec p'} 
= -\sum_{i,j=1}^N \sum_{\vec p, \vec p'} \frac{1}{2}V_{\vec p-\vec p'+\vec k_{ij}}n_{\vec p} n_{\vec p'}, \\
%\quad 
V_{\vec p-\vec p'} = \frac{2\pi e^2}{|\vec p-\vec p'|}, \quad 
n_{\vec p} = 1-\theta(|p|-p_0). \label{eq:exchange2}
\end{align}
%where $\vec k_{ij} = \vec k_{i}-\vec k_{j}$ is the momentum difference between two pocket centers. Exchange energy takes the form of a convolution.
where $\vec k_{ij} = \vec k_{i}- \vec k_{j}$ are momentum differences between pocket centers, $n_{\vec p}$ is the occupation number of the state with momentum $\vec p$ measured relative to the pocket center % of an arbitrary pocket
$
n_{\vec p} = 1-\theta(|p|-p_0)
$. Here $p_0$ is the radius of the circular Fermi surface %sea 
in each pocket 
\be
p_0 = \sqrt{4\pi n/N } .
\label{eq:p_0}
\ee
With these expressions, %we can proceed to evaluate 
the exchange energy can be evaluated analytically by performing the Fourier transform. 
Namely, perform Fourier transform:
\be
V(r) = \frac{e^2}{|r|}=\int \frac{d^2 p}{(2\pi)^2} e^{i\vec p \cdot\vec r} V_{\vec p}
\ee
and
\be
n(x) = \int \frac{d^2 p}{(2\pi)^2} e^{i \vec p \cdot \vec r} n_{\vec p} 
\ee
Then, the exchange energy can be written as
\begin{align}
&E_{ex}^{(N)} =- \frac{1}{2}\sum_{ij} \int d^2 r V(\vec r) n(\vec r)^2 e^{i \vec k_{ij}\cdot \vec r} \\ \nonumber
&= -\frac{1}{2}\sum_{ij} \int_0^\infty  V(r) n(r)^2 2\pi J_0 (r |\vec k_{ij}|)x dx
\end{align}
Here $J_0$ is the Bessel function.
To evaluate this quantity, we need to first work out the form of $n(r)$:
\begin{align}
n(r) = \int_{|p|<p_0}\frac{ dp_x dp_y }{4\pi^2}e^{ip_x r} %\\ \nonumber
= \int\frac{dp_x}{2\pi^2}  \sqrt{p_0^2-p_x^2} e^{ip_x r} 
.
\end{align}
Passing to polar coordinates, we have
\begin{align}
&n(r) = \frac{p_0^2}{2\pi^2}\int^{\frac{\pi}{2}}_{-\frac{\pi}{2}} d\theta \cos^2 \theta e^{i z \sin\theta} \\ \nonumber
&= \frac{p_0^2 }{8\pi} \lb J_{2}(z)+2J_{0}(z)+J_{-2}(z)\rb
= \frac{p_0^2 }{4\pi} \lb J_{2}(z)+J_{0}(z)\rb
\end{align}
where $z = rp_0$ , $J_0$ and $J_2$ are Bessel functions. With these expressions, we finally arrive at
%we can proceed to evaluate 
%the exchange energy can be evaluated analytically by performing the Fourier transform\cite{Supplement}, giving
\be
E_{ex}^{(N)} = -\frac{e^2 p_0^4}{16\pi} \sum_{ij} \int\limits_0^\infty dr \lb J_{2}(r p_0)+J_{0}(r p_0)\rb^2 J_0(r |\vec k_{ij}|). % \lp \frac{z|k_{ij}|}{p_0} \rp. 
\label{eq:E_ex_result}
\ee
%	\be
%	E_{ex}^{(N)} = -\frac{e^2 p_0^3}{16\pi} \sum_{ij} \int_0 dz \lb J_{2}(z)+J_{0}(z)\rb^2 J_0\lp \frac{z|\vec k_{ij}|}{p_0} \rp
%	\ee
Our isotropic parabolic bands model for pockets [see Eq.\eqref{eq:toy_model_1}] is expected to be accurate when the distance $k_*$ from the pocket centers to $K$ point is much greater than the pocket radius $p_0$. 
%The existence of three disconnected Fermi pockets requires $k_*\gg p_0$ where $k_*$ is the momentum of at the center of each pocket measured from $K$ point, and $p_0$ is the radius of each Fermi pocket. 
This yields an upper bound for carrier density: $n\lesssim 0.3 \times 10^{12}$ cm$^2$, where we used the value of $k_*$ 
%and $p_0$ 
estimated above. %in\cite{Supplement}. 
As Fig.~\ref{fig:PD} a) indicates, the maximal density in phase 1 always satisfies the above validity condition. We can therefore use the results in Eqs.~\eqref{eq:E_K_result},\eqref{eq:E_ex_result} to determine the phase diagram by comparing the  energies of one-pocket, two-pocket and three-pocket configurations.

\section{Momentum-polarized phases: observables and phenomenology} %Discussion}
%\addQ{In this section, we predict the observables that arise in momentum polarized phases.} 

There are several unique observables that can be predicted for the momentum polarized phases.
One surprising phenomenon that these phases display is the presence of persistent currents in the ground state, which are allowed by spontaneously-broken time reversal and inversion broken due to the transverse electric field. Such currents will not survive in a spatially uniform system bulk, yet they will show up at boundaries and interfaces. For example, they are expected to occur in the presence of spatial domains in which electrons populate different pockets. %a pair of counter flowing electric current along the boundary between two domains in which electrons populate different pockets. 
This behavior can be understood by parameterizing the momentum polarization using a position-dependent vector $\vec u(\vec x)$ as in the mean-field Hamiltonian given in Eq.\eqref{eq:uk}. %which is defined to be the momentum displacement of the center of Fermi pocket from the valley center. %This definition allows us to think of $\vec u(\vec x)$ as a fictitious gauge potential which couples to electrons through a momentum shift. 
%The momentum polarized phases can be described phenomenologically by an effective Hamiltonian Eq.\eqref{eq:uk}. 
In each domain, $\vec u(\vec x)$ is a uniform vector field aligned with a certain crystal axis. %Following from Bohr-Van Leeuwen theorem, 
For a uniform $\vec u$, the electric current in equilibrium equals zero %will vanish %inside each domain 
since the integral over the Fermi sea of carrier velocities derived from %the Hamiltonian 
Eq.~\eqref{eq:uk} will vanish. 
%Consequently, the orbital magnetization also vanishes in uniform system, although it is also allowed by the broken time reversal.%, similar to what happens when a uniform magnetic vector potential is applied. 
However, at a domain wall $\vec u(\vec x)$ varies in space, interpolates between different values in the domains. In this case a nonzero local current is allowed. This argument predicts a contribution to orbital magnetization first-order in spatial gradients of $\vec u(\vec x)$:
%are allowed due to %by the non-commutation relation between 
%the spatial-dependent $\vec u(\vec x)$ %and the electron momentum $k$
%. Performing a gradient expansion, at the leading order of $u$, we find the magnetization is given by 
	\be
	\vec m(\vec x) = \chi \vec \nabla\times \vec u(\vec x),\label{eq:m}
	\ee 
where the susceptibility $\chi$ is proportional to the Landau diamagnetic susceptibility. Here, terms such as $\vec \nabla \cdot \vec u$ must be excluded since $\vec m$ is an axial vector. Therefore, we expect a nonvanishing magnetization that peaks on the domain boundaries, originating from persistent currents that counter-propagate on the two sides of domain boundaries. Magnetization distribution that forms a network along domain boundaries is a directly testable signature of persistent currents. 
%This magnetization indicates a persistent current that counter-propagates on two sides of the domain wall.

Other interesting observables can arise from the broken crystallographic symmetries.
%Meanwhile, the momentum-polarized states further break the crystallographic symmetries, leading to distinct effects that can be probed by the symmetry-sensitive measurements discussed previously. 
Indeed, two possible orders of the ``parent" phase, i.e. phase 1, %which is the mother state of pocket polarization state, can take two forms, 
correspond to two kinds of broken symmetries--- either breaking only mirror symmetry, or breaking rotation, mirror and translation symmetries (see Table~\ref{table:symmetry}). If phase 1 only breaks mirror symmetry, then populating one or two out of three pockets will further break the three-fold rotation symmetry without breaking the translation symmetry, leading to electron nematicity. This symmetry breaking can be observed by measuring the anisotropy in the conductivity. If, however, the parent isospin order is intervalley coherent, then the only remaining symmetry to be broken in the pocket-polarization transition is the %discrete 
translation symmetry. Namely, the %easy-plane isospin order is a Kekul\'e charge density wave, which only break the lattice translation symmetry down to an commensurate translation symmetry. 
pocket polarization on top of valley coherent states transforms the Kekul\'e charge density wave into an incommensurate %Kekul\'e charge 
density wave which carries momentum  $\vec P_{i, j'} = 2 \vec K + \vec k_{i} + \vec k_{j'}$, $i,j=1,2,3$, $\vec k_{i'}=- \vec k_{i}$, see Fig.~\ref{fig:pocket_model} c). %, which breaks the commensurate translation down to an incommensurate translation symmetry. 
In this case, the pocket order can be detected by imaging long-period spatial modulations. 

%	the spontaneously broken time-reversal symmetry that gives rise to Hall conductivity also leads to a persistent current in the ground state, which has not been found in other graphene based systems. Similar to the Hall conductivity, the persistent current also changes abruptly at the onset of pocket orders.}

The momentum-polarized pocket orders can also be detected by measuring the Hall conductivity. %If the mother state is valley imbalance order $O_{1}^z$, the time reversal symmetry is broken, so the Hall conductivity is nonvanishing.
When pocket orders occur on top of the valley imbalance order $O_{1}^z$ which allows a nonvanishing Hall conductivity, the Hall conductivity changes abruptly since the Berry curvature is non-uniform near $K$ point [see Fig.~\ref{fig:pocket_model} d)].
% (Berry curvature in BLG: %$\Omega(p) \sim \frac{p^2}{D^2m^2}$, see \addZ{Supplement}).
% In comparison, %if the isospin ordering mother state takes the form of intervalley coherence 
If pocket ordering occurs on top of $O_{1}^{xy}$ isospin order, which originally respects the time reversal symmetry, enforcing %and enforces 
a vanishing Hall conductivity, then the onset of such a momentum polarization can break the time-reversal symmetry so long as electrons populate different pockets in valleys $K$ and $K'$ [e.g. pocket $1$ and $2'$ in Fig.~\ref{fig:pocket_model} c)]. As a result, the Hall conductivity will jump from zero to some finite value at the pocket ordering transition. Therefore, regardless of the form of the parent isospin order, we always expect a discontinuous behavior in Hall conductance at the onset of pocket orders.

	%Similarly, the change in Berry curvature at pocket polarizatioin transition can also be probed by measuring orbital magnetic moment.
	Another experimentally accessible signature of the Berry curvature %in this system 
is magnetization due to orbital currents in the ground state of the system. %orbital magnetization c moment. 
The magnetization can be estimated using the approach described in Refs.\cite{Xiao,Aryasetiawan}, %we estimate the orbital magnetic moment arising from Berry curvature to be 
giving $\sim 4$ Bohr magnetons per electron for the parameters used in Fig.~\ref{fig:pocket_model} d) [see Appendix \ref{sec:orbital magnetization} ]. This is few times larger than the orbital magnetic moments of electrons in a Landau level, and is readily measurable. 

In summary, exchange interactions in the flattened BBG bands result in a cascade of isospin-polarized orders and momentum-polarized orders. These orders are of interest for a number of reasons, in particular because they feature persistent currents and magnetization in the ground state.
% that we predict in BBG are a phenomenon of genuine interest because they feature  persistent currents and magnetization in the ground state. 
We stress that this phenomenon is distinct from orbital magnetization familiar in topological bands where it arises due to Berry phase. 
The momentum-polarized orders, rather than merely providing additional symmetry-breaking options by extending 4 isospin species to 12 isospin and pocket species, lead to unique physical properties such as nematic order with broken time reversal, persistent currents and Hall conductivity. Momentum polarization 
%Meanwhile, the unusual nematic order, which is very much unlike those nematic orders discussed in the literature, further adds significance to these phases – these orders break time-reversal symmetry and 
results in abrupt changes of the Berry curvature seen by electrons, leading to jumps in the anomalous Hall conductivity and orbital magnetization that can provide a convenient diagnostic of momentum polarization orders. 

This work was supported by the Science and Technology Center for
Integrated Quantum Materials, NSF Grant No. DMR1231319 and Army Research Office Grant W911NF-18-1-0116.

\appendix

%\section{Field-biased bilayer graphene bandstructure} 

\section{Estimating interaction strength $V$}\label{sec:estimate V}
In the main text, when numerically calculating the phase diagram, we are using the value of $V$ to represent the strength of exchange interaction. Here, we provide an  estimate for $V$ values.  

The interaction strength we used in our model in main text should correspond to the strength of the screened Coulomb interaction at the relevant momentum, which is Fermi momentum $p_0$, i.e.
\be
V = \tilde{V}_{p_0}.
\ee
Accounting for Thomas-Fermi screening, the screened Coulomb potential takes the following form
\be
\tilde{V}_{p_0} = \frac{V_{p_0}}{1+V_{p_0} \Pi_{p_0}},
\ee
where $\Pi_{p_0}$ is the polarization function at Fermi momentum. We estimate this quantity using the value of density of states at Fermi surface $\nu_0$. When the band is flat compared to the interaction energy, which is the case of our interest, we have
\be
\nu_0 V_{p_0} \gg 1
\ee
In this regime, the screened Coulomb interaction is approximately
\be
\tilde{V}_{p_0} = \frac{1}{\nu_0}.
\ee
Therefore, we can estimate the interaction as 
\be
V \sim \frac{1}{\nu_0} \sim 10^3 \text{ meV} \text{ nm}^2
\ee
where we have used $\nu_0 \sim n/W \sim 10^{-3}$ meV nm$^{-2}$, where $n$ is the carrier density $n\sim 10^{12}$ cm$^{-2}$, $W\sim 10$\,meV is the Fermi energy measured from the band bottom at %\addOO{said}
this 
carrier density. % $W\sim 10$ meV. 

%\section{A symmetry-based analysis of the isospin polarized states}

%\section{Three-pocket model}

\section{The Berry curvature and orbital magnetization} \label{sec:orbital magnetization}
It is straightforward to compute the Berry curvature using the Hamiltonian Eq.~\eqref{eq:Hmin}. 
Below, we first explain how we compute the Berry curvature in realistic BG model and obtain the result of Fig.\ref{fig:pocket_model}. We take the form of the Hamiltonian projected to conduction band in Eq.~\eqref{eq:Hmin}, and rewrite it as
\begin{align}
H_{3-p}^{min} &= \sum_{\vec p}\psi^\dagger_{i\vec p} \vec h(\vec p) \cdot \vec \tau_{ij} \psi_{j\vec p}, \\
%\quad 
&\vec h(p) \cdot \vec \tau = \left [h_0(\vec p) + h_t(\vec p)  + h_{D'}(\vec p)\right ]
\end{align}
where $\vec \tau = (\tau_1,\tau_2,\tau_3)$. Then the Berry curvature is given by
\be 
\Omega_{\vec p} = \frac{1}{2} \frac{\vec h}{|\vec h|} \cdot \lp \frac{\p \vec h(\vec p)}{\p p_1} \times \frac{\p \vec h(\vec p)}{\p p_2}\rp.\label{eq:sup Berry curvature}
\ee
In main text Fig.\ref{fig:pocket_model} we use Eq.\eqref{eq:sup Berry curvature} to numerically compute the Berry curvature.

Next, we estimate the orbital magnetization which arises from Berry curvature. Below, we recap the derivation of orbital moment described in Ref.~\onlinecite{Xiao,Aryasetiawan}, and apply it to our model. 

As a starting point, we consider the current flowing along the sample boundary, treating it as an anomalous current arising due to Berry's curvature and driven by the filed due to spatially varying trapping potential $U$. This gives a current value
\begin{align}
I &= e\int dx n(x)v(x) \\
&= e \int dx \int \frac{d^2p}{(2\pi)^2} \Omega_{\vec p} f(\epsilon_{\vec p} - \mu + U) \frac{\p U}{\p x}
\end{align}  
where $x$ is the coordinate in the direction perpendicular to the boundary.
The magnetization per unit area is therefore given by
\begin{align}
M &=  \frac{IA}{A} = e \int_{0}^{\tilde \mu} \Omega_{FS} (\tilde \mu-U) dU, \\%\quad 
&\Omega_{FS} (E) = \int \frac{d^2p}{(2\pi)^2} \Omega_{\vec p} f(\epsilon_{\vec p} - E) \label{eq:sup magnetization}
\end{align}
To estimate the magnetization value, we apply Eq.
\eqref{eq:sup magnetization} to the three-pocket model used in the main text,  taking $\Omega_p$ as a constant $\Omega_p \sim \Omega$ within the Fermi sea. This gives
\be
%\Omega_{FS} (E) = \frac{m_* E \Omega}{2\pi}, \quad
\frac{M}{\mu_B} \approx \frac{Nm_*m_{\rm e} \Omega\tilde{\mu}^2}{2\pi } .
\ee
where $\mu_B$ is the Bohr magneton, $m_{\rm e}$ is the electron mass, $N$ is the number of pockets that are filled, $\tilde{\mu}$ is Fermi level measured from the bottom of the band. We estimate $M$ for the case shown in Fig.\ref{fig:pocket_model}, where $N=3$, $\tilde{\mu}=10$ meV,  and $\Omega \sim 15$ nm$^2$ [extracted from Fig.~\ref{fig:pocket_model}], we find 
\be
\frac{M}{\mu_B} \sim 4\times 10^{-3} \mathrm{nm} ^{-2} \sim 4 n
\ee
where we used electron density $n=10^{11}$ cm$^{-2}$, a value corresponding to the regime where pocket polarization is expected. This predicts a sizable orbital magnetic moment of $\sim 4 $ Bohr magnetons per conduction electron.

\end{document}